\documentclass[showpacs,preprintnumbers,amsmath,amssymb,preprint,nofootinbib]{revtex4-1}
\usepackage{graphicx}
\usepackage{axodraw}
\usepackage{dcolumn}
\usepackage{hyperref}
\usepackage{bm}
\usepackage{amsmath,amssymb}

\def\be{\begin{equation}}
\def\ee{\end{equation}}
\def\bea{\begin{eqnarray}}
\def\eea{\end{eqnarray}}

\begin{document}

\title{Lepton-number-violating decays of heavy flavors induced by doubly-charged Higgs boson}
\vskip 6ex
\author{N\'estor Quintero}
\email{nquintero@fis.cinvestav.mx}
\affiliation{Departamento de F\'isica, Centro de Investigaci\'on y de Estudios Avanzados, Apartado Postal 14-740, 07000 M\'exico D.F., M\'exico}

\bigskip

\begin{abstract}
We study lepton-number-violating (LNV) decays of heavy flavors ($\tau$ lepton and top quark) induced by doubly-charged Higgs boson in the Higgs triplet model. It is found that the branching fractions of LNV $\tau^- \to \ell^{+}M_1^-M_2^-$ decays are highly suppressed compared with the current experimental limits. On the other hand, for LNV top quark decays, the most optimistic branching ratios for $t \to b\ell^{+}\ell^{+}W^{-}$ turn out to be at the level of $\sim 10^{-7} - 10^{-8}$. The observation of these rare top quark decays would be a clear signal of LNV processes, and their non-observation would allow us to constraint the parameters of the Higgs triplet model.
\end{abstract} 

\maketitle
\bigskip

\section{Introduction}
Neutrino oscillation experiments have established compelling evidence that neutrinos are massive \cite{PDG}. An explanation of the neutrino mass generation is one of the strongest motivations for considering new physics (NP) scenarios beyond the Standard Model (SM). Particularly, in the Higgs triplet model (HTM) \cite{THMmodels} neutrino masses can be generated as the product of leptonic Yukawa coupling ($h_{ij}$) and the Higgs triplet vacuum expectation value ($v_\Delta$). This is commonly known as the Type-II seesaw mechanism \cite{TypeII}. An important feature of the HTM is the existence of doubly-charged Higgs bosons ($H^{\pm\pm}$)\footnote{Doubly-charged Higgs bosons can also appear in many extensions of the SM with extended Higgs sectors, for instance, left-right symmetric models \cite{LRmodels}, 3-3-1 models \cite{331models}, and little Higgs models \cite{littlemodels}.}. They appear as a component of a $SU(2)_L$ scalar triplet and their couplings to lepton doublets violate lepton number conservation by two units. As a consequence they can induce lepton-number-violating (LNV) processes, as well as lepton-flavor-violating (LFV) ones. Therefore, a signal of a doubly-charged Higgs boson would be through the experimental observation of processes involving the production of same-sign dileptons in the final state. The discovery of $H^{\pm\pm}$ at current and future colliders would be an indicator of lepton number violation and also a signal of NP beyond the SM.

Theoretically, the production mechanism of doubly-charged Higgs bosons at the Large Hadron Collider (LHC) have been widely studied in the literature \cite{Akeroyd:2012,Sugiyama:2012,Melfo:2012,Aoki:2012,Chiang:2012,Chun:2012zu,Rentala,Akeroyd:2011,Akeroyd:2010,Garayoa,delAguila:2009,Akeroyd:2008,
Fileviez:2008,Akeroyd:2005,Chun}. Direct searches for $H^{\pm\pm}$ have been performed by several collaborations: OPAL \cite{OPAL}, L3 \cite{L3} and DELPHI \cite{DELPHI} at LEP; H1 \cite{H1} at HERA; CDF \cite{CDF} and D0 \cite{D0} at Tevatron; and also by CMS \cite{CMS} and ATLAS \cite{ATLAS:2012a,ATLAS:2012b} at the LHC. Up to date such particles have not yet been observed experimentally and their non-observation have provided strong constraints on their masses. Currently, the most stringent lower limits on their masses (excluded at 95\% confidence level) have been set by the CMS \cite{CMS} and ATLAS \cite{ATLAS:2012b} collaborations as is shown in Table \ref{masslimits}, by assuming that doubly-charged Higgs bosons decay mainly (100 \%) into same-sign dilepton channels with the same or different flavors, namely BR$(H^{\pm\pm} \to \ell_i^{\pm}\ell_j^{\pm}) = 1$.

\begin{table}[htb]
\label{masslimits} 
\caption{Current lower limits on the mass of $H^{\pm\pm}$ for different dilepton channels.}
\begin{tabular}{cccccccc}
\hline\hline 
 $\ell_i\ell_j$ & $ee$ & $e\mu$ & $\mu\mu$ & $e\tau$ & $\mu\tau$  & $\tau\tau$ & \\
 \hline
$m_{H^{\pm\pm}}$ (GeV) & 382 & 391  & 395 & 293 & 300 & 169 & CMS \cite{CMS}\\
 & 409 & 375 & 398 &  &  & & ATLAS \cite{ATLAS:2012b}\\
\hline\hline
\end{tabular}
\end{table}

Doubly-charged Higgs bosons can contribute to many low-energy LFV processes (see, for instance \cite{Chun,LFV1}), and also to LNV processes such as rare meson decays \cite{Picciotto:1997} and neutrinoless double beta ($0\nu\beta\beta$) decays \cite{neutrinoless}. Recently, contributions to the diphoton Higgs boson decay channel from loops of $H^{\pm\pm}$ have been also investigated \cite{Higgsdiphoton}. 



In this paper we study the possible effects of doubly-charged Higgs bosons in LNV decays of heavy flavors  $\tau^{-} \to \ell^+ M_1^{-}M_2^{-}$ ($\ell=e,\mu$ and $M_{1,2} = \pi, K$) and $t \to b\ell_i^{+}\ell_j^{+}W^{-}$ ($\ell_{i,j}=e,\mu,\tau$). These LNV $\tau$ decays have been previously discussed in the context of the exchange of a light (heavy) Majorana neutrino \cite{LNVtau,Atre:2005,Gribanov:2000,Atre:2009}, as well as in LNV top decays \cite{Quintero:2011,Eilam}. Here, we explore other underlying physics mechanism ($H^{\pm\pm}$) that could induce these LNV decays without involving Majorana neutrinos.

This work is organized as follows. In section \ref{HTM} we briefly review the general aspects of the
Higgs triplet model. We investigate and discuss the effects of doubly-charged Higgs bosons to LNV heavy flavor decays in section \ref{LNVflavor}. Our conclusions are presented in section \ref{Conclusion}.

\section{The Higgs triplet model}\label{HTM}
In this model the Higgs sector of the SM is extended by adding a $SU(2)_L$ Higgs triplet with hypercharge $Y=2$. The neutral component develops a vacuum expectation value (VEV) denoted by $v_\Delta$. The relevant LNV coupling to left-handed leptons is specified by the Yukawa interaction given by \cite{Akeroyd:2012,Melfo:2012,Aoki:2012}
\begin{equation}\label{Yukawa}
\mathcal{L}_Y^{\text{HTM}} = h_{ij} \psi_{iL}^{T} \ \mathcal{C}i\sigma_{2} \mathbf{\Delta} \psi_{jL} + h.c.\ ,
\end{equation}
where $\psi_{iL}^{T} = (\nu_{i} \ \ell_i)_{L}$ is the lepton doublet, $h_{ij}$ ($i,j=1,2,3$) are the entries of the $3\times 3$ leptonic Yukawa  coupling matrix, $\mathcal{C}$ is the charge conjugation operator, $\sigma_2$ is the Pauli matrix. The Higgs triplet in the $2\times 2$ matrix representation can be parametrized by
\begin{equation}
\mathbf{\Delta} = \left(
\begin{array}{cc}
\Delta^{+}/\sqrt{2} & \Delta^{++}\\
\Delta^{0} & -\Delta^{+}/\sqrt{2}\\
\end{array}
\right).
\end{equation} 
 
\noindent From expression (\ref{Yukawa}) it is clear that lepton number is explicitly broken by two units due to the leptonic Yukawa coupling $h_{ij}$. There are seven physical Higgs bosons in the HTM scheme: doubly-charged ($H^{\pm\pm}$), singly-charged ($H^{\pm}$), neutral $A^0$ (CP-odd), and neutral $h^0$ and $H^0$ (CP-even) \cite{Fileviez:2008}. The doubly-charged $H^{\pm\pm}$ is entirely composed of the triplet scalar field $\Delta^{\pm\pm}$. For the applications of the present
work we will consider the phenomenology related to the doubly-charged Higgs bosons. The doubly-charged Higgs bosons couple to the $W$ gauge bosons ($H^{\pm\pm}W^{\mp}_\mu W^{\mp}_\nu$) through the gauge coupling: $i\sqrt{2}g^{2}v_{\Delta} g_{\mu\nu}$ \cite{Fileviez:2008}.

After the neutral component Higgs triplet gets a VEV ($v_\Delta$), neutrinos acquire a Majorana mass $(m_\nu)_{ij} = \sqrt{2}h_{ij}v_{\Delta}$. In order to achieve a small neutrino mass less than 1 eV, $h_{ij} \sim 1$ and $v_\Delta \sim 1$ eV, or alternatively $h_{ij} \sim 10^{-10}$ and large $v_\Delta \sim 1$ GeV. This additional VEV $v_\Delta$ can be constrained from considering its effects on the $\rho$-parameter \cite{PDG}
\begin{equation}
\rho = M_{W}^2/M_{Z}^2 \cos^2\theta_W = \dfrac{1+2 v^{2}_\Delta/v^{2}}{1+4 v^{2}_\Delta/v^{2}},
\end{equation}

\noindent where $v= 246$ GeV is the VEV of the doublet Higgs field \cite{PDG}. By using current experimental values, one gets $v_\Delta/v \lesssim 0.01$, which constrains the triplet VEV to $v_\Delta \lesssim 3$ GeV.

Indirect experimental upper limits on the product of leptonic Yukawa couplings as a function of the mass $m_{H^{\pm\pm}}$ can be obtained from different processes mediated by a doubly-charged Higgs boson, such as: Bhabha scattering, LFV tri-leptonic and radiative decays of $\mu$ and $\tau$ leptons, muonium-antimuonium ($M\bar{M}$) conversion and $(g-2)_\mu$ measurement. Table \ref{hijlimits} shows the current limits on the products of couplings for various decay modes. So far, there is no limits on $h_{\tau\tau(\mu\tau)}^{2} / m_{H^{\pm\pm}}^{2}$, thus in this work, we will assume the same limit as for $h_{e\tau}^{2} / m_{H^{\pm\pm}}^{2}$.

\begin{table}\label{hijlimits} \small
\caption{\small Current upper bounds on the ratio of the product of leptonic Yukawa couplings and $m_{H^{\pm\pm}}^{2}$.}
\begin{tabular}{lc}
\hline\hline
Process & Limit (GeV$^{-2}$)\\
\hline 
Bhaba & \ \ \ $h_{ee}^{2} / m_{H^{\pm\pm}}^{2}$ $<$ 9.7 $\times 10^{-6}$ \cite{Swartz} \\
 & \ \ \ $h_{e\mu}^{2} / m_{H^{\pm\pm}}^{2}$ $<$ 1.0 $\times 10^{-6}$ \cite{Atag} \\
  & \ \ \ $h_{e\tau}^{2} / m_{H^{\pm\pm}}^{2}$ $<$ 1.0 $\times 10^{-6}$ \cite{Atag} \\
$\mu^{-} \to e^{+}e^{-}e^{-}$ & \ \ \ $h_{ee}h_{e\mu}/m_{H^{\pm\pm}}^{2}$ $<$ 4.7 $\times 10^{-11}$ \cite{Rentala} \\
$\tau^{-} \to e^{+}e^{-}e^{-}$ & \ \ \ $h_{ee}h_{e\tau}/m_{H^{\pm\pm}}^{2}$ $<$ 2.09 $\times 10^{-8}$ \cite{Rentala} \\
$\tau^{-} \to \mu^{+}e^{-}e^{-}$ & \ \ \ $h_{ee}h_{\mu\tau}/m_{H^{\pm\pm}}^{2}$ $<$ 1.56 $\times 10^{-8}$ \cite{Rentala} \\
$\tau^{-} \to e^{+}e^{-}\mu^{-}$ & \ \ \ $h_{e\mu}h_{e\tau}/m_{H^{\pm\pm}}^{2}$ $<$ 2.57 $\times 10^{-8}$ \cite{Rentala} \\
$\tau^{-} \to \mu^{+}\mu^{-}e^{-}$ & \ \ \ $h_{e\mu}h_{e\tau}/m_{H^{\pm\pm}}^{2}$ $<$ 3.0 $\times 10^{-8}$ \cite{Rentala} \\
$\tau^{-} \to \mu^{+}\mu^{-}\mu^{-}$ & \ \ \ $h_{\mu\mu}h_{\mu\tau}/m_{H^{\pm\pm}}^{2}$ $<$ 1.97 $\times 10^{-8}$ \cite{Rentala} \\
$\mu^{-} \to e^{-}\gamma$ & \ \ \ $(2 h_{ee}h_{e\mu} + 2 h_{e\mu}h_{\mu\mu}+h_{e\tau}h_{\mu\tau})/m_{H^{\pm\pm}}^{2}$ $<$ 5.8 $\times 10^{-9}$ \cite{Rentala} \\
$\tau^{-} \to e^{-}\gamma$ & \ \ \  $(2 h_{ee}h_{e\tau} + 2 h_{e\tau}h_{\tau\tau}+h_{e\mu}h_{\mu\tau})/m_{H^{\pm\pm}}^{2}$ $<$ 7.2 $\times 10^{-7}$ \cite{Rentala}\\
$\tau^{-} \to \mu^{-}\gamma$ & \ \ \ $(h_{e\mu}h_{e\tau} + 2 h_{\mu\mu}h_{\mu\tau}+ 2 h_{\mu\tau}h_{\tau\tau})/m_{H^{\pm\pm}}^{2}$ $<$ 8.3 $\times 10^{-7}$ \cite{Rentala}\\
$M\bar{M}$ conversion & \ \ \ $h_{ee}h_{\mu\mu}/m_{H^{\pm\pm}}^{2}$ $<$ 1.98 $\times 10^{-7}$ \cite{Rentala}\\
$(g-2)_\mu$ & \ \ \ $h_{\mu\mu}^{2} / m_{H^{\pm\pm}}^{2}$ $<$ 2.5 $\times 10^{-5}$ \cite{Swartz}\\
\hline\hline
\end{tabular}
\end{table}

\section{LNV heavy flavor decays induced by a doubly-charged Higgs boson}\label{LNVflavor}
In this section we carry out the calculation of LNV decays of $\tau$ lepton and top quark decays, mediated by a doubly-charged Higgs boson in the context of HTM previously discussed. Similar studies about the effects of $H^{\pm\pm}$ in LNV meson decays were done in Ref. \cite{Picciotto:1997}.​

\begin{figure}
\centering
\includegraphics[scale=0.5]{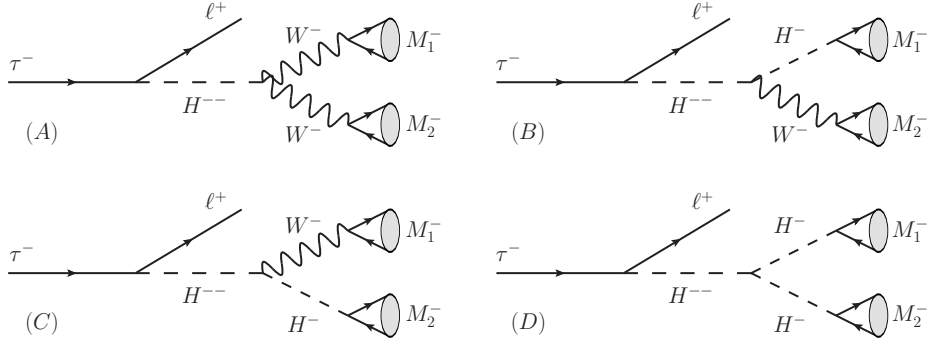}
\caption{\small Feynman diagrams mediated by a doubly- and singly-charged Higgs boson contributing to LNV $\tau^{-} \to \ell^{+}M_1^{-}M_2^{-}$ decays.}
\label{Fig:LNVtau}
\end{figure}

\subsection{LNV $\tau$ decays}\label{Sec.LNVtau}
Let us first consider the LNV $\tau^{-}(p) \to \ell^{+}(p_1)M_1^{-}(p_2)M_2^{-}(p_3)$ decays ($\ell=e, \mu$). All possible diagrams contributing to this process are shown in Fig. \ref{Fig:LNVtau}. The decay amplitude  corresponding to the diagram in Fig. \ref{Fig:LNVtau}(A) is given by
\begin{eqnarray}
\mathcal{M}_A^{\tau} = && \ L_{PS}h_{\ell \tau} \Big(\dfrac{\sqrt{2}g^{2}v_{\Delta}}{m_{H^{--}}^{2}}\Big)  \Big(\dfrac{ig}{2\sqrt{2}}\Big)^{2} \Big(\dfrac{V_{M_1}^{\text{CKM}}V_{M_2}^{\text{CKM}}}{m_W^{4}}\Big) f_{M_1}f_{M_2} \ N_{M_1 M_2} (p_2 \cdot p_3), 
\end{eqnarray}

\noindent where $V_{M_1}^{\text{CKM}} (V_{M_2}^{\text{CKM}})$ and $f_{M_1} (f_{M_2})$ are the Cabibbo-Kobayashi-Maskawa (CKM) matrix element and decay constant for the meson $M_1 (M_2)$; $L_{PS} = [\bar{u}^{c}_{\ell}(p_1)P_L u_{\tau}(p)]$ is the pseudoscalar leptonic current. The factor $N_{M_1 M_2} = 1 (2)$ for different (identical) mesons in the final state.
This process can also receive contribution from diagram \ref{Fig:LNVtau}(B), however a rather conservative of the ratio of the amplitudes (B)/(A) gives 
\begin{eqnarray}
\dfrac{\mathcal{M}_B^{\tau}}{\mathcal{M}_A^{\tau}} &\simeq &  \dfrac{(m_u - m_{q_i})}{(m_u + m_{q_i})}\Big(\dfrac{m_W}{m_{H^-}}\Big)^{2}\Big(\dfrac{1}{V_{M_1}^{\text{CKM}}}\Big)  \Big(\dfrac{m_{M_1}}{gv}\Big)^{2},  \nonumber\\
&\lesssim &  10^{-7}, 
\end{eqnarray}

\noindent where we have used typical values of $(m_u - m_{d})/(m_u + m_{d}) = 1/3$, $m_{M_1} \sim 0.1$ GeV, $V_{M_1}^{\text{CKM}} \sim 1$ and for the singly-charged Higgs boson mass $m_{H^-} > 90$ GeV \cite{ATLAS:Charged}. Thus, the contribution from Fig. \ref{Fig:LNVtau}(B) is very small. Similarly, \ref{Fig:LNVtau}(C) and \ref{Fig:LNVtau}(D) are also suppressed. Therefore, we keep the diagram shown in Fig. \ref{Fig:LNVtau}(A) as the dominant contribution.

After performing the spin-averaged squared amplitude the decay width is parametrized as follows \cite{PDG}
\begin{equation}
\Gamma(\tau^{-} \to \ell^{+}M_1^{-}M_2^{-}) = \Big(1 - \dfrac{1}{2}\delta_{M_{1}M_{2}} \Big) \dfrac{1}{(2\pi)^{3}32 m_{\tau}^{3}}  \int_{s_{12}^{-}}^{s_{12}^{+}}  ds_{12} \int_{s_{23}^{-}}^{s_{23}^{+}} ds_{23} \ |\overline{\mathcal{M}_A^{\tau}}|^{2},
\end{equation}

\noindent where $s_{12}=(p_1 + p_2)^{2}$ and $s_{23}=(p_2 + p_3)^{2}$ are the invariant mass variables. The factor $(1-\delta_{M_1 M_2}/2)$ accounts for identical mesons in the final state. The integration limits are given by
\begin{eqnarray}
s_{23}^{\pm}(s_{12}) =&& m_{M_1}^{2} + m_{M_2}^{2} + \dfrac{1}{2 s_{12}} \Big[(m_{\tau}^{2} - m_{M_2}^{2}- s_{12})(s_{12} - m_{\ell}^{2} + m_{M_1}^{2}) \pm \sqrt{\lambda(s_{12},m_{\ell}^{2},m_{M_1}^{2})}  \nonumber\\
&& \times \sqrt{\lambda(m_{\tau}^{2},s_{12},m_{M_2}^{2})} \Big],
\end{eqnarray}
\begin{equation}
s_{12}^{-} = (m_{\ell} + m_{M_1})^{2}; \ \ s_{12}^{+} = (m_{\tau} - m_{M_2})^{2},
\end{equation}

\begin{table}
\caption{\label{table:resultTHM}  Branching ratios (BR) of LNV $\tau$ lepton decays induced by a doubly-charged Higgs boson.}
\begin{tabular}{lcc}
\hline\hline 
 Decay modes &  BR &  Exp. limits (90 \% C.L.) \cite{Belle} \\
\hline
 $\tau^- \to e^{+}\pi^-\pi^-$&  $ < 4.66 \times 10^{-23} $ &  $ < 2.0 \times 10^{-8}$ \\
 $\tau^- \to e^{+}\pi^-K^-$ &  $ < 4.14 \times 10^{-24} $ &  $ < 3.2 \times 10^{-8}$ \\
 $\tau^- \to e^{+}K^-K^-$ &  $ < 8.06 \times 10^{-26} $ &  $ < 3.3 \times 10^{-8}$ \\
 $\tau^- \to \mu^{+}\pi^-\pi^-$ &  $ < 4.41 \times 10^{-23} $ &  $ < 3.9 \times 10^{-8}$ \\
 $\tau^- \to \mu^{+}\pi^-K^-$ &  $ < 3.87 \times 10^{-24} $ &  $ < 4.8 \times 10^{-8}$ \\
 $\tau^- \to \mu^{+}K^-K^-$ &  $ < 7.34 \times 10^{-26} $ &  $ < 4.7 \times 10^{-8}$ \\
\hline\hline
\end{tabular}
\end{table}

\noindent with $\lambda(x,y,z)=x^{2}+y^{2}+z^{2}-2xy-2xz-2yz$. Taking a value of $v_{\Delta} \lesssim$ 3 GeV, Yukawa couplings $h_{e\tau}^{2} (h_{\mu\tau}^{2})$ from Table \ref{hijlimits}, and adopting the lower limits on the mass $m_{H^{--}}$ from CMS Collaboration (Table \ref{masslimits}), we display our numerical results on the branching ratios in Table \ref{table:resultTHM}. We get that branching ratios induced by doubly-charged Higgs boson are highly suppressed, compared with current experimental limits. These results are consistent with similar studies on LNV meson decays \cite{Picciotto:1997}; in both cases the amplitudes are strongly suppressed by four powers of the weak coupling and inverse powers of  $m_{H^{--}}$ and $m_{W}$. 

The $\tau^- \to \ell^{+}M_1^-M_2^-$ decays induced by the exchange of Majorana neutrino have been previously studied  \cite{LNVtau,Atre:2005,Gribanov:2000,Atre:2009}. These decays can be strongly enhanced for neutrino masses in the range $m_{M_2} + m_{\ell} \leq m_N \leq m_{\tau} - m_{M_1}$, which allows to get significant constrains on the mixings of Majorana and active neutrinos by  using experimental bounds \cite{Gribanov:2000,Atre:2009}. For neutrino masses outside of this interval the rates of these channels become extremely small, even as the ones listed in Table \ref{table:resultTHM}.

\begin{figure}
\centering
\includegraphics[scale=0.5]{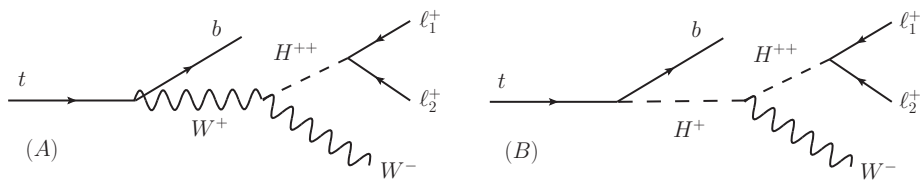}
\caption{\small Feynman diagrams mediated by a doubly- and singly-charged Higgs boson contributing to LNV decays $t \to b\ell_i^{+}\ell_j^{+}W^{-}$.}
\label{Fig:LNVtop}
\end{figure}

\subsection{LNV top decays}
We consider now LNV $t(p) \to b(p_1)\ell_i^{+}(p_2)\ell_j^{+}(p_3)W^{-}(p_4)$ decays ($\ell_{i,j}=e,\mu,\tau$) induced by a doubly-charged Higgs boson $H^{++}$ (see Fig. \ref{Fig:LNVtop}). By a similar reasoning as in section \ref{Sec.LNVtau}, we discard the effects of a singly-charged Higgs boson, and we will take the diagram \ref{Fig:LNVtop}(A) as the dominant contribution. The decay amplitude associated to diagram \ref{Fig:LNVtop}(A) is given by

\begin{eqnarray}\label{Amptop}
\mathcal{M}_A^{\text{top}} = &&  \Big(\dfrac{ig}{2\sqrt{2}}\Big) V_{tb}
[\bar{u}_{b}(p_1) \gamma_{\mu}(1-\gamma_5)u_{t}(p)] \Delta_{W}^{\mu\nu}(Q) \Big(\dfrac{\sqrt{2}g^{2}v_{\Delta} g_{\nu\alpha}}{(p_2 + p_3)^{2} - m_{H^{++}}^{2}}\Big)  \ L_{PS}h_{ij} \ \varepsilon_{W}^{\alpha},
\end{eqnarray}

\noindent where
\begin{equation}
\Delta_W^{\mu\nu}(Q)=i(-g^{\mu\nu}+Q^{\mu}Q^{\nu}/m_W^2)/(Q^2-m_W^2+im_W\Gamma_{W}), 
\end{equation}

\noindent with $Q=p-p_1$, denotes the $W$ boson propagator in the unitary gauge, $\varepsilon_W^{\alpha}$ is the four-vector polarization of the $W$ boson, and  $L_{PS} = [\bar{u}^{c}_{i}(p_2)P_L u_{j}(p_3) - (p_2\rightleftharpoons p_3)]$.

Following the definitions given in Ref. \cite{Quintero:2011,Lopez:2005}, the kinematics of four-body decays can be described in terms of five independent variables $\{s_{12}, s_{34}, \theta_1, \theta_3, \phi\}$. With this choice of kinematics, the decay rate can be written as
\begin{eqnarray}\label{Widthtop}
\Gamma^{\text{top}}_{\ell_i\ell_j} &\equiv & \Gamma(t \to b\ell_i^{+}\ell_j^{+} W^{-}) \nonumber\\
&=& \Big(1 - \frac{\delta_{\ell_i\ell_j}}{2} \Big) \frac{X\beta_{12}\beta_{34}}{4(4\pi)^6 m_{t}^3} |\overline{\mathcal{M}_A^{\text{top}}}|^{2} \cdot d\Phi \ ,
\end{eqnarray}

\noindent with $d\Phi = ds_{12}ds_{34}d\cos \theta_1 d\cos \theta_3 d\phi$ the phase space factor, given in terms of
 $s_{12}=(p_1+p_2)^2$ and $s_{34}=(p_3+p_4)^2$ the invariant masses of the 12 and 34 particles, and angular variables $(\theta_1,\theta_3, \phi)$ \cite{Quintero:2011,Lopez:2005}. Identical leptons in the final state phase-space are taken into account through the factor $(1-\delta_{\ell_i\ell_j}/2)$. $|\overline{\mathcal{M}_A^{\text{top}}}|^{2}$ is the spin-averaged squared amplitude, $\beta_{12}$ ($\beta_{34}$) is the velocity of particle 1 (particle 3) in the center of mass frame of particles 1 and 2 (3 and 4) and $X\equiv [(p^2-s_{12}-s_{34})^2-4s_{12}s_{34}]^{1/2}$.

\begin{figure}[htbp]
\includegraphics[scale=0.41]{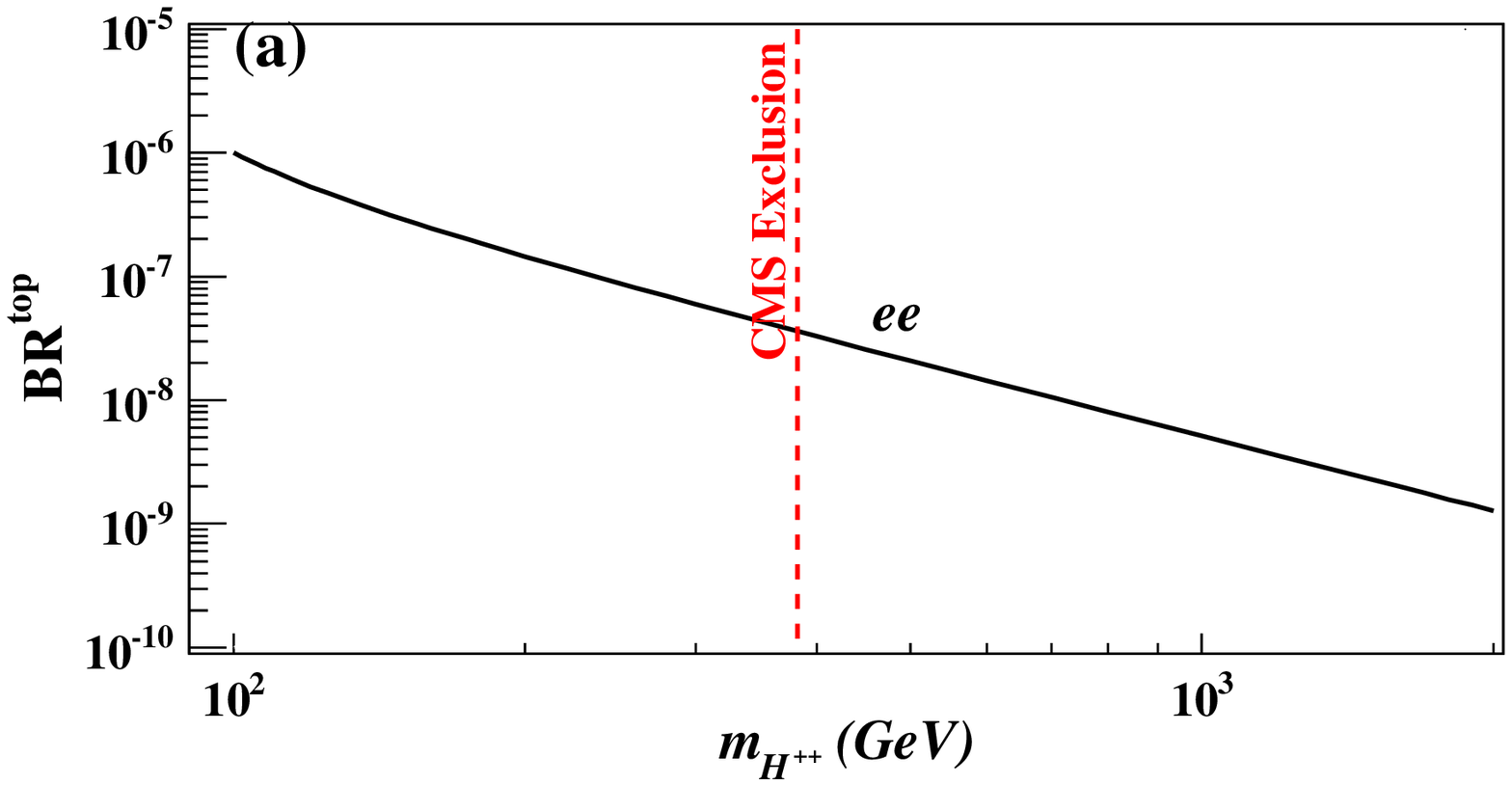}\hfill
\includegraphics[scale=0.41]{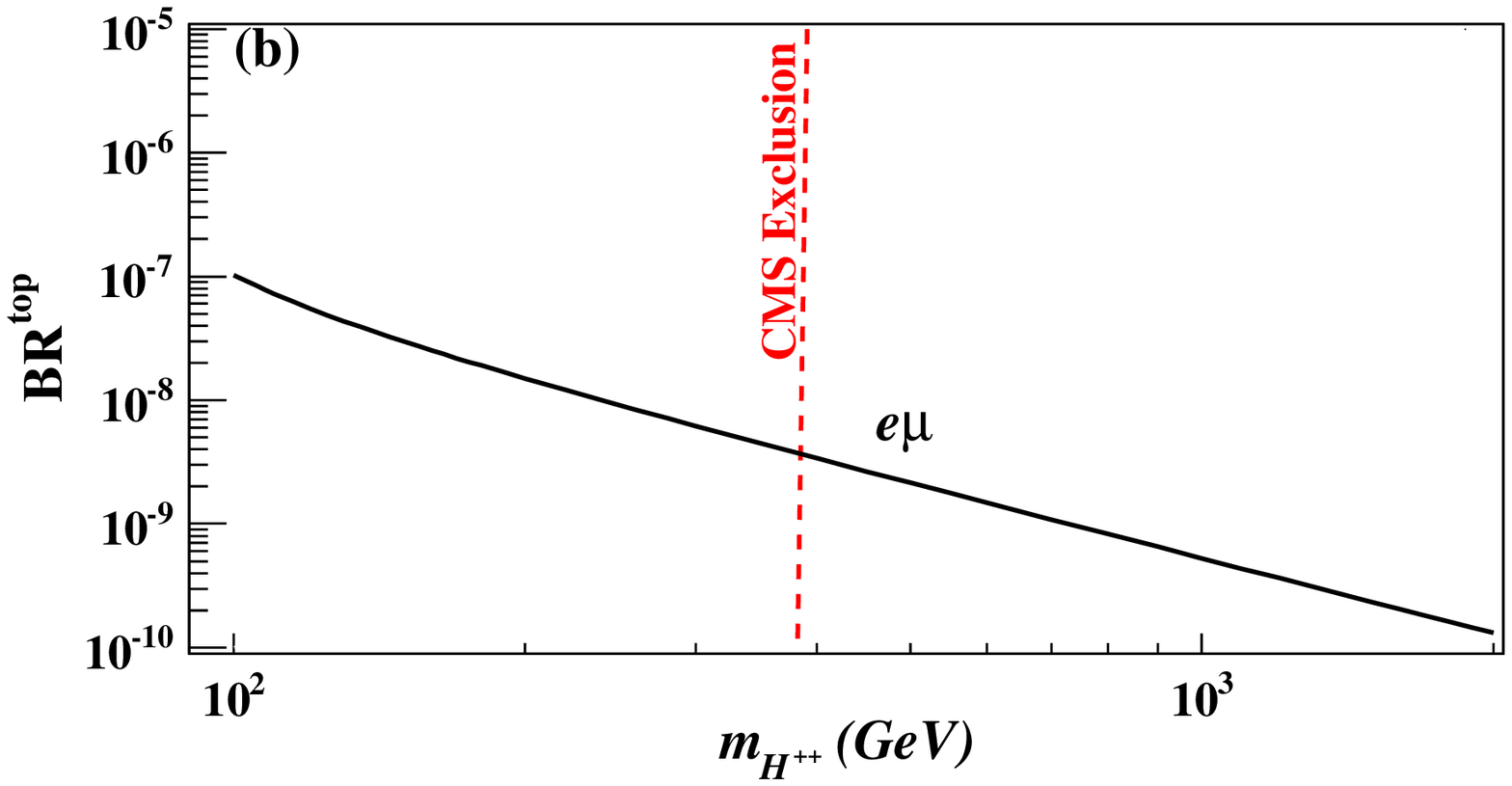} \\
\includegraphics[scale=0.41]{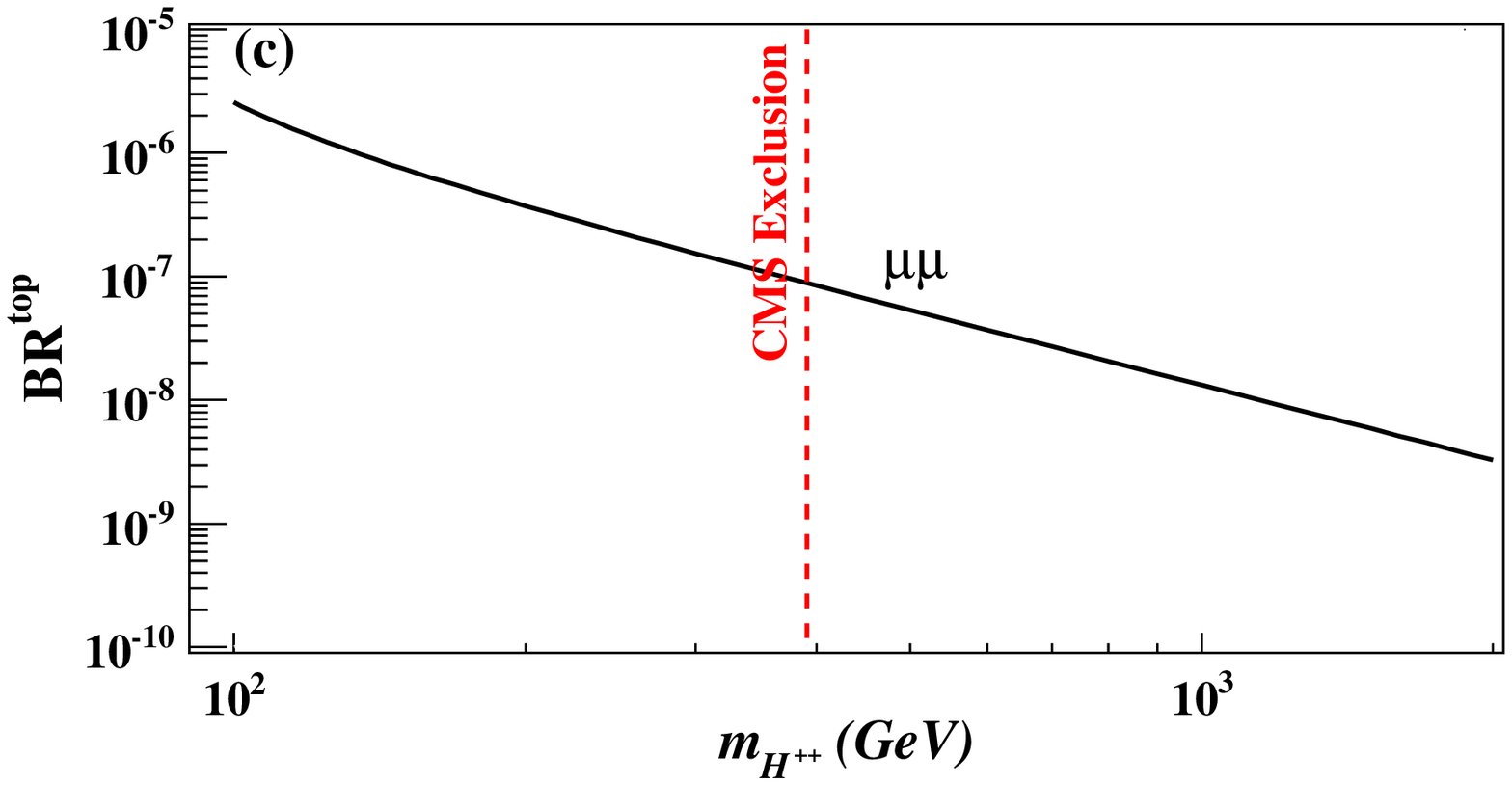}\hfill
\includegraphics[scale=0.41]{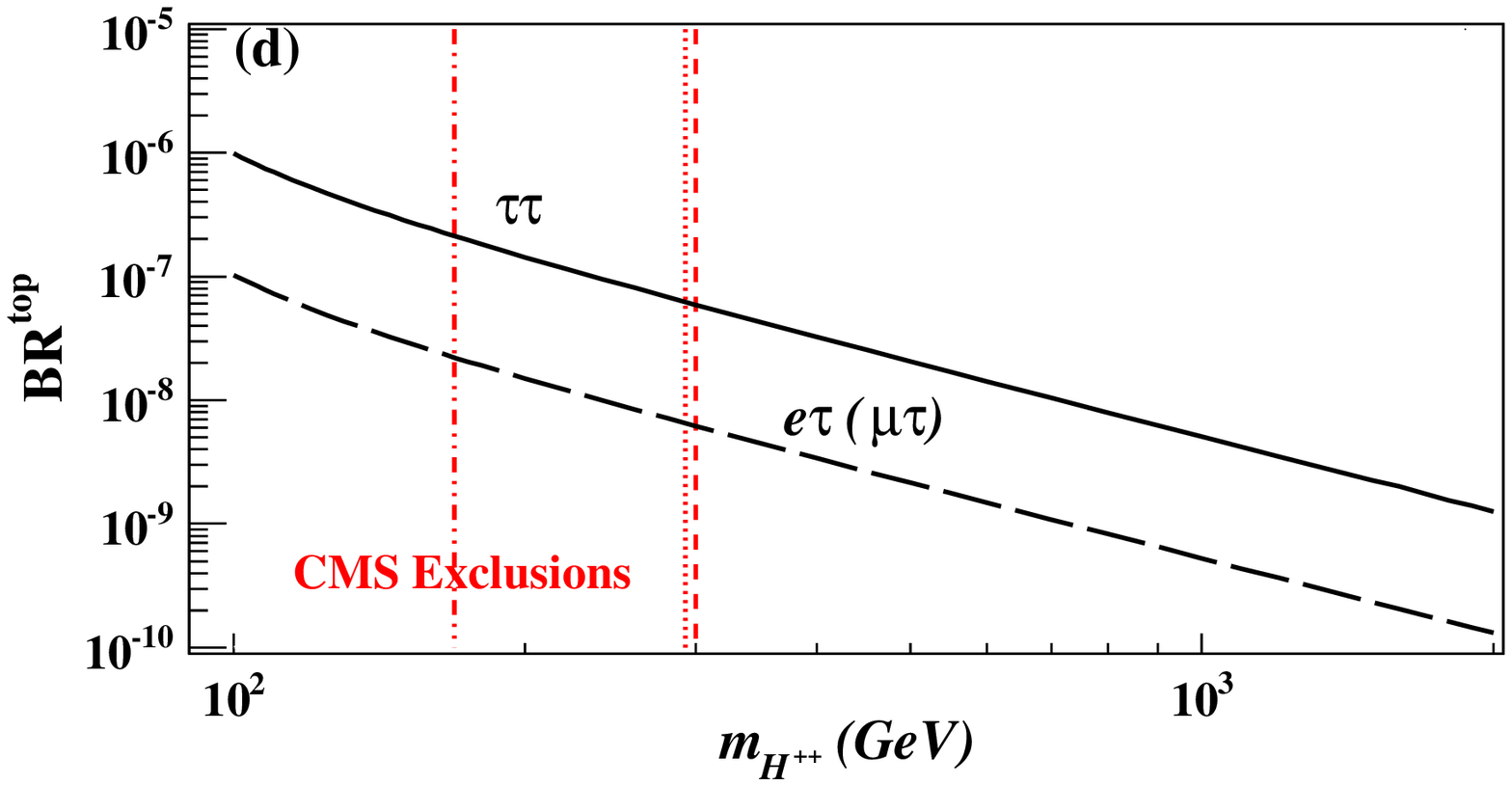} \\
\caption{\small Branching ratio of LNV top decays as a function of the $H^{++}$ mass for exclusive decays to different same-sign dilepton channels: (a) $ee$, (b) $e\mu$, (c) $\mu\mu$ [solid line] and (d) $e\tau,\mu\tau$ [long-dashed line], $\tau\tau$ [solid line]. The vertical lines correspond to the respective limits on $m_{H^{++}}$ from CMS (Table \ref{masslimits}).}
\label{fig:BRtop} 
\end{figure}

In order to illustrate the effects of doubly-charged Higgs boson in LNV $t \to b\ell_i^{+}\ell_j^{+}W^{-}$ decays, in Fig. \ref{fig:BRtop}  we plot the branching ratios (BR$^{\text{top}}_{\ell_i\ell_j}$) for different dilepton channels as a function of the doubly-charged Higgs boson mass $m_{H^{++}}$ in the range of 100 GeV up to 2 TeV. These plots were obtained by using a value of $v_{\Delta} \lesssim$ 3 GeV and taking the relevant products of leptonic Yukawa couplings from Table \ref{hijlimits}. For the sake of simplicity we include only the limits on $m_{H^{++}}$ from CMS collaboration (Table \ref{masslimits}) since they do not differ too much from ATLAS ones.
As we can observe from Fig. \ref{fig:BRtop}, the maximum (as well as optimistic) values for the branching ratios are the following: (a) BR$^{\text{top}}_{ee} \lesssim 10^{-8}$ for $m_{H^{++}} \simeq 420$ GeV, (b) BR$^{\text{top}}_{e\mu} \lesssim 10^{-9}$ for $m_{H^{++}} \simeq 400$ GeV, (c) BR$^{\text{top}}_{\mu\mu} \lesssim 10^{-7}$ for $m_{H^{++}} \simeq 400$ GeV, and (d) BR$^{\text{top}}_{e\tau,\mu\tau} \lesssim 10^{-9}$ for $m_{H^{++}} \simeq 310$ GeV and BR$^{\text{top}}_{\tau\tau} \lesssim 10^{-7}$ for $m_{H^{++}} \simeq 180$ GeV. Let us notice that these BR$^{\text{top}}$ are of the same order of magnitude as the ones induced by virtual heavy Majorana neutrinos \cite{Quintero:2011,Eilam}. 
So far, no experimental searches have been reported for these LNV top quark decays. The LHC can explore a large variety of top quark physics \cite{topreview}, however, currently there is not enough sensitivity to test our predictions. Eventually, when LHC luminosity is increased (Super-LHC) \cite{Gianotti}, one can expect that these branching ratios could be within the LHC's reach. Indeed, it is expected that at 10 TeV the LHC will be a top quark factory \cite{topLHC}. 

\begin{figure}
\centering
\includegraphics[scale=0.5]{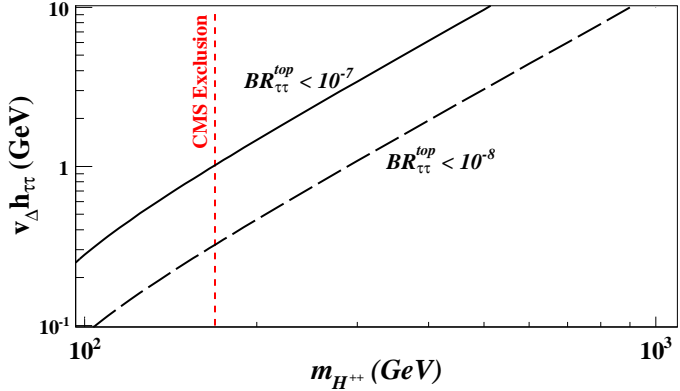}
\caption{\small The product of $v_\Delta h_{\tau\tau}$ parameters as function of the mass $m_{H^{++}}$, by assuming BR$^{\text{top}}_{\tau\tau} < 10^{-7}$ (solid line) and $10^{-8}$ (dashed line). The vertical line corresponds to limit on $m_{H^{++}}$ from CMS (Table \ref{masslimits})}
\label{Fig:vDelta}
\end{figure}

When upper limits on the branching ratios of LNV $t \to b\ell_i^{+}\ell_j^{+}W^{-}$ decays become available, one would be able to get constraints on the product of $v_\Delta h_{ij}$ parameters (without any assumption on the individual parameter) as a function of the mass $m_{H^{++}}$. In the case of the $ee$-channel the experimental limits on $0\nu\beta\beta$ decays of nuclei can provide stronger constraints on this product. Indeed, a comparison of the decay amplitudes for this decay mode in the framework of the HTM and the one due to the exchange of light Majorana neutrinos (standard mechanism) leads to $(v_\Delta h_{ee}/m_{H^{--}}^2) \sim \langle m_{\beta\beta} \rangle /\langle q\rangle ^2$ where $\langle q\rangle \sim $ 100 MeV is the typical momentum of the virtual Majorana neutrino. By using a value of effective Majorana mass of $\langle m_{\beta\beta} \rangle \lesssim 10^{-1}$ eV, we get a stringent limit $(v_\Delta h_{ee}/m_{H^{--}}^2 ) \lesssim 10^{-8}$ GeV$^{-1}$. If we use this limit, we get a strong constraint of BR$^{\text{top}}_{ee} \lesssim 10^{-11}$ for $m_{H^{++}} \simeq 420$ GeV. This clearly indicates a smaller branching fraction than the one obtained in Fig. \ref{fig:BRtop}(a). Nevertheless, in the case of other same-sign channels ($e\mu, e\tau, \mu\mu, \mu\tau, \tau\tau$), LNV top quark decays are capable to provide information on the other parameters (not yet constrained) as is shown in Fig. \ref{Fig:vDelta}, for instance, for the case of the $\tau\tau$ channel by assuming BR$^{\text{top}}_{\tau\tau} < 10^{-7}$ (solid line) and $10^{-8}$ (dashed line). 
Additionally, since the leptonic Yukawa coupling $h_{ij}$ is related to the neutrino mass matrix through $v_\Delta$, as it was discussed in Sec. \ref{HTM}, this would allow us to obtain constraints on the relative magnitude of each element of the neutrino mass matrix as a function of the mass $m_{H^{++}}$.

\section{Conclusions}\label{Conclusion}
In this work we have studied LNV decays of heavy flavors ($\tau$ lepton and top quark)  induced by a doubly-charged Higgs boson, in the context of a Higgs triplet model. Using current bounds on relevant couplings, we have found that for LNV $\tau$ lepton decays the corresponding branching ratios turns out to be very small and far below the sensitivities of current and future experiments, leading to unobservable rates. For the case of LNV top quark decays, we obtained branching ratios around $\sim 10^{-7} - 10^{-8}$ in the most optimistic case. The top quark system may be a good place to probe this NP scenario, and the non-observation of these LNV decays would allow us to constraint the product of $v_\Delta h_{ij}$ parameters of the HTM.
 
\begin{acknowledgments}
The author would like to thank Conacyt (M\'exico) for financial support, and G. L\'opez Castro for a careful reading of the manuscript, useful discussions and comments. He is also grateful to P. Langacker for helpful suggestions.

\end{acknowledgments}



\begin{thebibliography}{99}

\bibitem{PDG} 
J. Beringer \textit{et al.} [Particle Data Group], Phys. Rev. D \textbf{86}, 010001 (2012).

\bibitem{THMmodels}
G. B. Gelmini, M. Roncadelli, Phys. Lett. B \textbf{99}, 411 (1981);
H. Georgi, M. Machacek, Nucl. Phys. \textbf{B262}, 463 (1985);
J. F. Gunion, R. Vega, and J. Wudka, Phys. Rev. D \textbf{42}, 1673 (1990). 

\bibitem{TypeII}
W. Konetschny and W. Kummer, Phys. Lett. B \textbf{70}, 433 (1977); 
T. P. Cheng and L. F. Li, Phys. Rev. D \textbf{22}, 2860 (1980); 
M. Magg and C. Wetterich, Phys. Lett. B \textbf{94}, 61 (1980); 
J. Schechter and J. W. F. Valle, Phys. Rev. D \textbf{22}, 2227 (1980); 
R. N. Mohapatra and G. Senjanovi\'c, Phys. Rev. D \textbf{23}, 165 (1981);
G. Lazarides, Q. Shafi, and C. Wetterich, Nucl. Phys. \textbf{B181}, 287 (1981).

\bibitem{LRmodels}
J. C. Pati, A. Salam, Phys. Rev. D \textbf{10},  275 (1975), Erratum-\textit{ibid.} D \textbf{11}, 703 (1975);
R. N. Mohapatra, J. C. Pati, Phys. Rev. D \textbf{11}, 566 (1975);
G. Senjanovi\'c, R. N. Mohapatra, Phys. Rev. D \textbf{12}, 1502 (1975);
T. G. Rizzo, Phys. Rev. D \textbf{25},  1355 (1982).

\bibitem{331models}
F. Pisano, V. Pleitez, Phys. Rev. D \textbf{46}, 410 (1992); 	
P. H. Frampton, Phys. Rev. Lett. \textbf{69}, 2889 (1992);
J. E. Cieza Montalvo, N. V. Cortez, J. Sa Borges, and M. D. Tonasse, Nucl. Phys. \textbf{B756}, 1 (2006);
\textbf{B796}, 422(E) (2008). 


\bibitem{littlemodels}
For recent review see: M. Schmaltz, D. Tucker-Smith, Annu. Rev. Nucl. Parti. Sci. \textbf{55}, 229 (2005);
T. Han, H. E. Logan, and L. T. Wang, JHEP \textbf{01}, 099 (2006).

\bibitem{Akeroyd:2012}
A. G. Akeroyd, S. Moretti, and H. Sugiyama, Phys. Rev. D \textbf{85}, 055026 (2012).

\bibitem{Sugiyama:2012}
H. Sugiyama, K. Tsumura, and H. Yokoya, Phys. Lett. B \textbf{717}, 229 (2012).

\bibitem{Melfo:2012} 
A. Melfo, M. Nemevsek, F. Nesti, G. Senjanovi\'c, and Y. Zhang, Phys. Rev. D \textbf{85},  055018 (2012).

\bibitem{Aoki:2012}
M. Aoki, S. Kanemura and K. Yagyu, Phys. Rev. D \textbf{85}, 055007 (2012).

\bibitem{Chiang:2012}
C. W. Chiang, T. Nomura and K. Tsumura, Phys. Rev. D \textbf{85}, 095023 (2012).

\bibitem{Chun:2012zu} 
E.~J.~Chun and P.~Sharma,
  JHEP {\bf 1208}, 162 (2012).
  
\bibitem{Rentala}
V. Rentala, W. Shepherd, and S. Su, Phys. Rev. D \textbf{84}, 035004 (2011).

\bibitem{Akeroyd:2011}
A. G. Akeroyd and S. Moretti, Phys. Rev. D \textbf{84}, 035028 (2011).

\bibitem{Akeroyd:2010}
A. G. Akeroyd, C. W. Chiang, and N. Gaur, JHEP \textbf{11}, 005 (2010).

\bibitem{delAguila:2009}  
F. del Aguila and J. A. Aguilar-Saavedra, Nucl. Phys. \textbf{B813}, 22 (2009).

\bibitem{Garayoa}
J. Garayoa and T. Schwetz, JHEP \textbf{03}, 009 (2008).
	 
\bibitem{Akeroyd:2008}
A. G. Akeroyd, M. Aoki, and H. Sugiyama, Phys. Rev. D \textbf{77}, 075010 (2008). 

\bibitem{Fileviez:2008}
P. Fileviez P\'erez, T. Han, G.-y. Huang, T. Li and K. Wang, Phys. Rev. D \textbf{78}, 015018 (2008); 

\bibitem{Akeroyd:2005}
M. Kadastik, M. Raidal and L. Rebane, Phys. Rev. D \textbf{77}, 115023 (2008);
A. G. Akeroyd and M. Aoki, Phys. Rev. D \textbf{72}, 035011 (2005);
E. Ma, M. Raidal and U. Sarkar, Nucl. Phys. \textbf{B615}, 313 (2001).
\bibitem{Chun}
E. J. Chun, K. Y. Lee and S. C. Park, Phys. Lett. B \textbf{566}, 142 (2003). 

\bibitem{OPAL}
P. D. Acton \textit{et al}. [OPAL Collaboration], Phys. Lett. B \textbf{295}, 347 (1992);
G. Abbiendi \textit{et al}. [OPAL Collaboration], Phys. Lett. B \textbf{526}, 221 (2002);
\textbf{577}, 93 (2003).

\bibitem{L3}
P. Achard \textit{et al}. [L3 Collaboration], Phys. Lett. B \textbf{576}, 18 (2003).

\bibitem{DELPHI}
J. Abdallah  \textit{et al}. [DELPHI Collaboration], Phys. Lett. B \textbf{552}, 127 (2003).

\bibitem{H1}
A. Aktas \textit{et al}. [H1 Collaboration], Phys. Lett. B \textbf{638}, 432 (2006).

\bibitem{CDF}
D. E. Acosta \textit{et al}. [CDF Collaboration], Phys. Rev. Lett. \textbf{93}, 221802 (2004); 
T. Aaltonen \textit{et al}. [CDF Collaboration], Phys. Rev. Lett. \textbf{101}, 121801 (2008);
\textbf{107}, 181801 (2011).

\bibitem{D0}
V. M. Abazov \textit{et al}. [D0 Collaboration], Phys. Rev. Lett. \textbf{101}, 071803 (2008);
\textbf{108}, 021801 (2012).

\bibitem{CMS}
S. Chatrchyan \textit{et al}. [CMS Collaboration], Eur. Phys. J. C \textbf{72}, 2189 (2012).

\bibitem{ATLAS:2012a}
G. Aad \textit{et al}. [ATLAS Collaboration], Phys. Rev. D \textbf{85}, 032004 (2012).

\bibitem{ATLAS:2012b}
G. Aad \textit{et al}. [ATLAS Collaboration], Eur. Phys. J. C \textbf{72} 2244 (2012).  


\bibitem{LFV1}
T. Fukuyama, H. Sugiyama, and K. Tsumura, JHEP \textbf{03}, 044 (2010);
A. G. Akeroyd, M. Aoki, and H. Sugiyama, Phys. Rev. D \textbf{79}, 113010 (2009);
A. Abada, C. Biggio, F. Bonnet, M. B. Gavel, and T. Hambye, JHEP \textbf{03}, 061 (2007).

\bibitem{Picciotto:1997}
C. Picciotto, Phys. Rev. D \textbf{56}, 1612 (1997); Y.-L. Ma, Phys. Rev. D \textbf{79}, 033014 (2009).

\bibitem{neutrinoless}
R. N. Mohapatra and J. D. Vergados, Phys. Rev. Lett. \textbf{47}, 1713 (1981).
J. Schechter and J. W. F. Valle, Phys. Rev. D \textbf{25}, 2951 (1982);
L. Wolfenstein, Phys. Rev. D \textbf{26}, 2507 (1982);
W. C. Haxton, S. P. Rosen, and G. J. Stephenson, Phys. Rev. D \textbf{26}, 1805 (1982);
H. V. Klapdor-Kleingrothaus and U. Sarkar, Phys. Lett. B \textbf{554}, 45 (2003);
  
\bibitem{Higgsdiphoton}
A. G. Akeroyd and S. Moretti, Phys. Rev. D \textbf{86}, 035015 (2012);
A. Arhrib, R. Benbrik, M. Chabab, G. Moultakae, and L. Rahilib, JHEP \textbf{04}, 136 (2012);
I. Picek and B. Radovci\'c, arXiv:1210.6449  [hep-ph]; 
L. Wang, X.-F. Han, arXiv:1209.0376 [hep-ph]; Phys. Rev. D \textbf{86}, 095007 (2012).


\bibitem{LNVtau}
A. Ilakovac, B. A. Kniehl, and A. Pilaftsis, Phys. Rev. D \textbf{52}, 3993 (1995); 
A. Ilakovac and A. Pilaftsis, Nucl. Phys. \textbf{B437}, 491 (1995); 
A. Ilakovac, Phys. Rev. D \textbf{54}, 5653 (1996); 

\bibitem{Atre:2005}
A.~Atre, V.~Barger and T.~Han, 
Phys. Rev. D {\bf 71}, 113014 (2005). 

\bibitem{Gribanov:2000}
V. Gribanov, S. Kovalenko, and I. Schmidt,	
Nucl. Phys. \textbf{B607}, 355 (2001).
      
\bibitem{Atre:2009}
A. Atre, T. Han, S. Pascoli, and B. Zhang, JHEP \textbf{05}, 030 (2009).

\bibitem{Quintero:2011}
D. Delepine, G. L\'opez Castro, and N. Quintero,  Phys. Rev. D \textbf{84}, 096011 (2011) [Erratum-\textit{ibid} D \textbf{86}, 079905 (2012)].

\bibitem{Eilam}
S. Bar-Shalom, N. G. Deshpande, G. Eilam, J. Jiang, and A. Soni, Phys. Lett. B \textbf{643}, 342 (2006).

\bibitem{Swartz}
M. L. Swartz, Phys. Rev. D \textbf{40}, 1521 (1989).

\bibitem{Atag}
S. Atag and K. O. Ozansoy, Phys. Rev. D \textbf{70}, 053001 (2004).

\bibitem{ATLAS:Charged}
ATLAS Collaboration., ATLAS-CONF-2011-151; ATLAS-COM-CONF-2011-174.

\bibitem{Belle}
Y. Miyazaki {\it et al.} [Belle Collaboration], arXiv:1206.5595 [hep-ex].

\bibitem{Lopez:2005}
A. Flores-Tlalpa, G. L\'opez Castro, and G. Toledo Sanchez, Phys. Rev. D \textbf{72}, 113003 (2005).

\bibitem{topreview}
W. Bernreuther, J. Phys. G \textbf{35}, 083001 (2008).

\bibitem{Gianotti}
F. Gianotti \textit{et al}., Eur. Phys. J. C \textbf{39}, 293 (2005).

\bibitem{topLHC}
S. Mehlhase, \textit{Prospects for Top Physics at the start-up LHC}, ATL-PHYS-SLIDE-2009-275.

\end{thebibliography}
\end{document}